\documentclass{emulateapj}\usepackage{apjfonts}
\newcommand{\rxj}{RX~J1856.5$-$3754}
\newcommand{\cxo}{{\em Chandra}}
\newcommand{\chandra}{{\em Chandra}}
\newcommand{\xmm}{{\em XMM}}
\defcitealias{kvk05}{KvK05a}
\defcitealias{kvk05b}{KvK05b}
\defcitealias{tm07}{TM07}

\newcommand{\expnt}[2]{\ensuremath{#1 \times 10^{#2}}}   

\begin{document}

\title{Timing the Nearby Isolated Neutron Star \rxj}

\author{M.~H.~van Kerkwijk\altaffilmark{1} and D.~L.~Kaplan\altaffilmark{2}}

\altaffiltext{1}{Department of Astronomy and Astrophysics, University
  of Toronto, 50 St.\ George Street, Toronto, ON M5S 3H4, Canada;
  mhvk@astro.utoronto.ca}

\altaffiltext{2}{Hubble Fellow; MIT Kavli Institute for Astrophysics and Space
  Research, Massachusetts Institute of Technology, 77 Massachusetts
  Avenue, 37-664H, Cambridge, MA 02139, USA; dlk@space.mit.edu}

\slugcomment{Accepted for publication in ApJ (Letters), 18 Dec.\ 2008}

\begin{abstract}
\rxj\ is the X-ray brightest among the nearby isolated neutron stars.
Its X-ray spectrum is thermal, and is reproduced remarkably well by a
black-body, but its interpretation has remained puzzling.  One reason
is that the source did not exhibit pulsations, and hence a magnetic
field strength---vital input to atmosphere models---could not be
estimated.  Recently, however, very weak pulsations were discovered.
Here, we analyze these in detail, using all available data from the
{\em XMM-Newton} and {\em Chandra} X-ray observatories.  From
frequency measurements, we set a 2$\sigma$ upper limit to the
frequency derivative of $|\dot\nu|<
1.3\times10^{-14}{\rm\,Hz\,s^{-1}}$.  Trying possible phase-connected
timing solutions, we find that one solution is far more likely than
the others, and we infer a most probable value of $\dot\nu =
(-5.98\pm0.14)\times 10^{-16}{\rm\,Hz\,s^{-1}}$.  The inferred
magnetic field strength is $1.5\times10^{13}{\rm\,G}$, comparable to
what was found for similar neutron stars.  From models, the field
seems too strong to be consistent with the absence of spectral
features for non-condensed atmospheres.  It is sufficiently strong,
however, that the surface could be condensed, but only if it is
consists of heavy elements like iron.  Our measurements imply a
characteristic age of $\sim\!4\,$Myr.  This is longer than the cooling
and kinematic ages, as was found for similar objects, but at almost a
factor ten, the discrepancy is more extreme.  A puzzle raised by our
measurement is that the implied rotational energy loss rate of
$\sim\!\expnt{3}{30}{\rm\,erg\,s}^{-1}$ is orders of magnitude smaller
than what was inferred from the H$\alpha$ nebula surrounding the
source.
\end{abstract}


\section{Introduction}
\label{sec:intro}
The nearby neutron star \rxj\ (hereafter J1856) is the closest and
brightest of the seven radio-quiet, isolated neutron stars (INS)
discovered by \textit{ROSAT} (for reviews, see \citealt{haberl07} and
\citealt{vkk07}).  The INSs have attracted much attention, in part
because of the hope that the equation of state in their ultradense
interiors could be constrained using their thermal emission.  Progress
has been slow, however, as adequate fits of the spectra appear to
require somewhat contrived models, with the most successful one
appealing to hydrogen layers of finely tuned thickness superposed on a
condensed surface \citep[e.g.,][]{mzh03,hkc+07}.  A major hindrance is
our ignorance of the surface magnetic field: without an observational
constraint, models can consider too wide a range to be useful.

We have been able to make some progress, using X-ray observations to
determine coherent timing solutions and hence estimates of the dipolar
magnetic fields for two of the INS \citep[][ hereafter
  \citetalias{kvk05},b]{kvk05,kvk05b}.  However, J1856 has resisted
such attempts, as its pulsations were so weak that they were
discovered only recently, in a long \xmm\ observation \citep[][
  hereafter \citetalias{tm07}]{tm07}.  \citetalias{tm07} were unable
to determine the spin-down rate of the source, finding only a limit of
$|\dot\nu| < 4\times10^{-14}{\rm\,Hz\,s^{-1}}$ (at 90\% confidence).
Here, from a detailed analysis of a larger amount of data, we infer a
stronger constraint, and derive a most likely phase-connected
solution.
     
\begin{deluxetable*}{llrrrrlll}
\tablewidth{0.75\hsize}
\tablecaption{Log of Observations And Timing Measurements\label{tab:obs}}
\tablehead{&& 
\colhead{$\Delta T$}&
&
\colhead{$f_{\rm b}$}&
&
\colhead{$a$}&
\colhead{TOA}& 
\colhead{$\nu$}\\
\colhead{Date}&\colhead{Instr.\tablenotemark{a}}&
\colhead{(ks)}&
\colhead{$N_{\rm ev}$}&
\colhead{(\%)}&
\colhead{$Z_1^2$}&
\colhead{(\%)}&
\colhead{(MJD)}&
\colhead{(Hz)}
}
\startdata
\dataset[ADS/Sa.CXO#113]{2000 Mar 10}\dotfill & S    
   &  55 &  36799 &  1 &\nodata &\nodata &\nodata &\nodata\\ 
\dataset[ADS/Sa.CXO#3380,ADS/Sa.CXO#3381,ADS/Sa.CXO#3382,ADS/Sa.CXO#3399]{2001 Oct 8--15}$\ldots$ & S    
   & 626 & 320541 &  2 & 13 & 0.9(2)  & 52193.786169(4)  & 0.1417396(3) \\ 
\dataset[ADS/Sa.XMM#0106260101]{2002 Apr 8}\dotfill & Xm1ti
   &  58 & 664279 &  1 & 34 & 1.01(17)& 52373.011797(2)  & 0.1417403(17)\\ 
\dataset[ADS/Sa.CXO#4286]{2002 Aug 6}\dotfill & I    
   &  10 &   9984 &  9 &\nodata &\nodata &\nodata &\nodata\\ 
\dataset[ADS/Sa.CXO#4287]{2002 Sep 3}\dotfill & I    
   &  55 &  55632 & 11 &\nodata &\nodata &\nodata &\nodata\\ 
\dataset[ADS/Sa.CXO#4288]{2002 Sep 23}\dotfill & Ifoc 
   &   8 &  14627 &    &\nodata &\nodata &\nodata &\nodata\\ 
\dataset[ADS/Sa.CXO#4356]{2003 May 4}\dotfill & I    
   &  50 &  51544 &  9 &\nodata &\nodata &\nodata &\nodata\\ 
\dataset[ADS/Sa.XMM#0201590101]{2004 Apr 17}\dotfill & Xpnti
   &  65 & 377313 &  9 & 18 & 1.0(2)  & 53113.309266(3)  & 0.141741(3)  \\ 
\dataset[ADS/Sa.XMM#0165971601,ADS/Sa.XMM#0165971701]{2004 Sep 24}\dotfill & Xmmf 
   &  71 & 463180 &    & 21 & 1.0(2)  & 53272.338862(3)  & 0.1417401(19)\\ 
\dataset[ADS/Sa.CXO#5174]{2004 Nov 4}\dotfill & I    
   &  49 &  47443 & 14 &\nodata &\nodata &\nodata &\nodata\\ 
\dataset[ADS/Sa.XMM#0165971901]{2005 Mar 23}\dotfill & Xm1lw
   &  35 & 378678 &    & 17 & 0.9(2)  & 53452.558481(3)  & 0.141738(4)  \\ 
\dataset[ADS/Sa.XMM#0213080101]{2005 Apr 15}\dotfill & Xnopn
   &   9 &  28357 &    &\nodata &\nodata &\nodata &\nodata\\ 
\dataset[ADS/Sa.CXO#6094]{2005 Jun 10}\dotfill & I    
   &  50 &  65608 &  2 &\nodata &\nodata &\nodata &\nodata\\ 
\dataset[ADS/Sa.XMM#0165972001]{2005 Sep 24}\dotfill & X    
   &  35 & 366372 &    & 29 & 1.3(2)  & 53637.530095(3)  & 0.141740(3)  \\ 
\dataset[ADS/Sa.CXO#6095]{2005 Nov 13}\dotfill & I    
   &  48 &  35586 & 37 &\nodata &\nodata &\nodata &\nodata\\ 
\dataset[ADS/Sa.XMM#0165972101]{2006 Mar 26}\dotfill & X    
   &  70 & 772404 &    & 16 & 0.65(16)& 53821.055080(3)  & 0.1417375(15)\\ 
\dataset[ADS/Sa.CXO#7052]{2006 Jun 1}\dotfill & I    
   &  46 &  33270 & 40 &\nodata &\nodata &\nodata &\nodata\\ 
\dataset[ADS/Sa.XMM#0412600101]{2006 Oct 24}\dotfill & X    
   &  73 & 794684 &    & 52 & 1.14(16)& 54032.442099(2)  & 0.1417412(11)\\ 
\dataset[ADS/Sa.CXO#7053]{2006 Nov 18}\dotfill & I    
   &  50 &  31997 & 40 &\nodata &\nodata &\nodata &\nodata\\ 
\dataset[ADS/Sa.XMM#0412600201]{2007 Mar 14}\dotfill & Xodf 
   &  69 & 734710 &    & 55 & 1.22(17)& 54174.246834(2)  & 0.1417357(12)\\ 
\dataset[ADS/Sa.XMM#0415180101]{2007 Mar 25}\dotfill & Xmlw 
   &  41 & 488457 &    & 31 & 1.1(2)  & 54184.467408(3)  & 0.141739(2)  \\ 
\dataset[ADS/Sa.XMM#0412600301]{2007 Oct 4}\dotfill & X    
   &  70 & 416891 &    & 11 & 0.7(2)  & 54377.621398(4)  & 0.1417405(19)
\enddata
\tablecomments{Explanation of columns: (date) date of the observation;
  (ins) instrument code as given in note (a); ($\Delta T$) time
  between first and last event; ($N_{\rm ev}$) number of events
  extracted; ($f_{\rm b}$) estimated fraction of events due to
  background (given only if $\geq\!1$\%); ($Z_1^2$) power of the
  pulsations; ($a$) fractional amplitude; (TOA) time of maximum light
  closest to the mean of all event times; ($\nu$) frequency.
  Power and timing measurements were derived only for observations
  with $N_{\rm ev}>\expnt{2}{5}$.  All uncertainties are $1\sigma$;
  those on TOA and $\nu$ were derived from fits with $a$ fixed to
  $0.96(1-f_{\rm b})$\% (see text).}
\tablenotetext{a}{Instrument codes are: (S) \chandra\ High-Resolution
  Camera for Spectroscopy, used with the Low-Energy Transmission
  Grating; (I) \chandra\ High-Resolution Camera for Imaging, out of
  focus; (Ifoc) as (I), but in focus; (X) \xmm\ European Photon
  Imaging Cameras, combining data from the PN and MOS detectors using
  small-window mode and thin filter; (Xm1ti) as (X), but with MOS1
  used in timing mode (for which $f_b=8$\%); (Xpnti) PN used in timing
  mode, and both MOS used in full-frame mode and thus ignored; (Xmmf)
  for MOS, a portion was taken using the medium filters; (Xm1lw) MOS1
  used in large-window mode; (Xnopn) only the MOS were used; (Xodf)
  duplicate frames in the observation data files were removed; (Xmlw)
  both MOS used in large-window mode.}
\end{deluxetable*}

\section{Observations}
\label{sec:obs}
We retrieved all observations of J1856 taken with the {\em XMM-Newton}
(\xmm) and {\em Chandra} X-ray Observatories (see
Table~\ref{tab:obs}), and reprocessed the data from \xmm\/'s European
Photon Imaging Cameras with PN and MOS detectors using the {\tt
  epchain} and {\tt emchain} pipelines in SAS version 7.1, and those
from \cxo's High Resolution Cameras following standard threads in CIAO
version 4.0.  For the \xmm\ data from 2007 Mar 14, the pipelines gave
warnings about odd time jumps in both MOS and PN.  We traced these to
small sets of duplicated events in the observation data files (ODF),
which we removed (our results do not depend on this, or on whether or
not we include this observation).

Given that the pulsations are so weak, we tried to optimize the number
of source counts.  For the PN imaging observations, we decreased the
default 150\,eV low-energy cutoff to 100\,eV, which leads to a 50\%
increase in source events.  For both MOS and PN, we selected source
events from a circular region of $37\farcs5$ radius and with energies
below 1\,keV (where background flares are negligible).  Following
normal practice, we included only one and two-pixel events (patterns 0
to 4) with no warning flags for PN, and single to triple events
(patterns 0 to 12) with the default flag mask for MOS (removing
$<\!0.1$\% of the source counts).

There are also two \xmm\ timing observations.  For the one with MOS1,
we used default settings, and selected single to triple events from
columns 315 to 329
with energies less than 1\,keV and the default flag mask.  The timing
observation with PN suffers from frequent bursts of low-energy noise
events, and from some high-energy ones due to flares.  We identified
the former using {\tt epreject}, after which the source dominates the
background down to 215\,eV for single-pixel events and down to 430\,eV
for doubles, and up to 600\,eV during flares and up to 800\,eV
otherwise.  We selected source events using these criteria from
columns 29 to~45.

For the \cxo\ spectra, we extracted not only the zeroth order source
events, as done by \citetalias{tm07}, but also the about two times
more numerous diffracted events.  For the former, we used a circle
with radius of $3\farcs6$, while for the latter we took rectangles in
transmission grating angles $({\rm tg_r},{\rm tg_d})$ defined by
$|{\rm tg_d}|<0\fdg000531$ (the default for extracting spectra), and
$0\fdg12<|{\rm tg_r}|<0\fdg42$ (corresponding to
$21\lesssim\lambda\lesssim73$\,\AA), where the source clearly
dominates the background.

Most of the \chandra\ imaging observations were not taken in focus,
and we used ellipses to define regions where the source clearly
dominated the background.  We extracted events with pulse intensity in
the range 1--220 only, since few source events had higher values.

\begin{figure}
\includegraphics[width=\hsize]{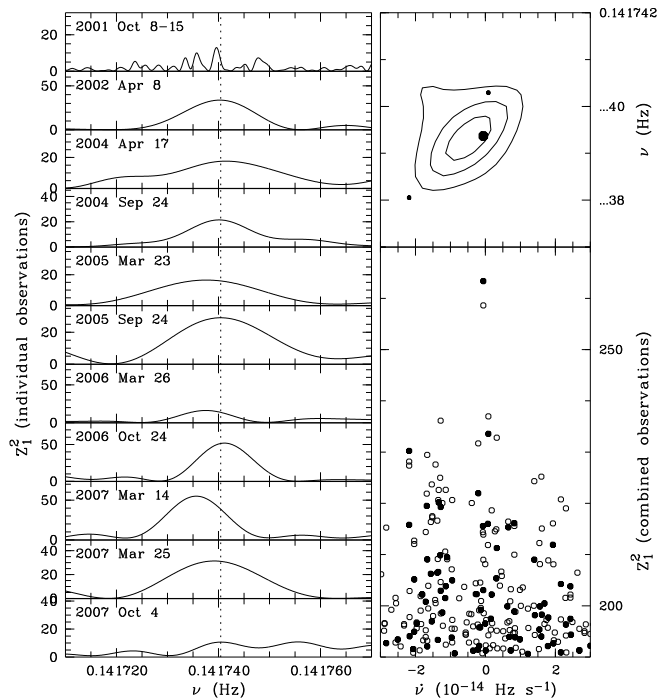}
\caption{Power spectra for \rxj.  {\em(left)} Results for the long
  individual data sets, near the pulse frequency identified by
  \citetalias{tm07} (dotted line).  The ordinate is scaled such that
  the mid-point corresponds to a fractional pulsation amplitude
  $a=(2Z_1^2/N_{\rm ev})^{1/2}=1$\%. {\em(top right)} Incoherently
  combined power spectra for a range of period derivatives.  The
  contours are at $\Delta Z_1^2=2.3$, 6.17, 11.8 (or 1, 2, and
  3$\sigma$ for two parameters of interest).  The circles represent
  the three highest peaks from the coherent analysis.  {\em(bottom
    right)} Power for coherent combinations of the long observations
  (open circles) and all observations (filled circles).\label{fig:ft}
}
\end{figure}

\section{Incoherent Analysis}
\label{sec:incoherent}
Given the fractional amplitude of only $a\simeq1$\% of the pulsations
in J1856, a 3-$\sigma$ detection requires
$\sim\!2(3/a)^2\simeq2\times10^5$ counts (for a single trial at a
known frequency).  Combining events from the PN and MOS cameras, ten
out of eleven \xmm\ observations have sufficient counts, while only
the long spectrum suffices among the \cxo\ observations.  For all of
these, we computed $Z_1^2$ power spectra \citep{bbb+83} for the
barycentered event times in a narrow interval around the frequency
found by \citetalias{tm07}.  As can be seen in Fig.~\ref{fig:ft}, the
pulsations are detected in all eleven observations.  We then fitted
the event times for each observation with a sinusoid using
\citet{cash79} minimization.  The resulting fractional amplitudes $a$,
frequencies $\nu$, and inferred arrival times TOA are listed in
Table~\ref{tab:obs}.

One sees that the different observations give similar amplitudes.
Correcting them for background and statistical bias (the expectation
value is $\langle a_{\rm m}\rangle=([a_0(1-f_{\rm b})]^2+4/N_{\rm
  ev})^{1/2}$, where $a_{\rm m}$ and $a_0$ are the measured and
unbiased amplitudes, $N_{\rm ev}$ the number of counts, and $f_{\rm
  b}$ the fraction due to background), and taking a weighted average,
the best estimate of the amplitude is~$0.96\pm0.06$\% ($\chi^2=12.2$
for 10 degrees of freedom [DOF]).

To obtain the best measurements of the phases and frequencies, we
refitted each observation holding the amplitude fixed at $a_{\rm
  i}=0.96(1-f_{\rm b})\%$.  This yielded the same best-fit values,
but uncertainties that were slightly increased for observations for
which $a_{\rm m}$ was high, and decreased for those for which $a_{\rm
  m}$ was low.  Analytically, this is expected: for small amplitudes,
$a$ is not covariant with $\nu$ or $\phi$, and the uncertainties scale
with $(a_{\rm m}/a_{\rm i})^{-1/2}$.  In Table~\ref{tab:obs}, we list
uncertainties from this second fit.

Fitting the frequencies gives a best-fit frequency derivative
$\dot\nu=(-5\pm4)\times10^{-15}{\rm\,Hz\,s^{-1}}$, with $\chi^2=13.7$
for 9 DOF (see Table~\ref{tab:solution}).  Clearly, one cannot exclude
a constant frequency (which has $\chi^2=15.0$ for 10 DOF).  The
2-$\sigma$ upper limit is
$|\dot\nu|<1.3\times10^{-14}{\rm\,Hz\,s^{-1}}$ (a factor of three
improved over \citetalias{tm07}, mostly due to the precise
\cxo\ frequency).  To verify our result, we also added the $Z_1^2$
power spectra for a range of frequency derivatives; this led to the
same best-fit values, and showed no significant other peaks (see
Fig.~\ref{fig:ft}).

\section{Coherent Analysis}
\label{sec:coherent}

Since J1856 has been observed so often, we attempted a coherent
analysis, using two different methods.  First, we tried finding
$(\nu,\dot\nu)$ combinations that were consistent with both the
frequencies and the arrival times listed in Table~\ref{tab:obs}.  For
this purpose, we generalised the method of \citetalias{kvk05} to allow
us to explore many possible cycle counts between observations (no pair
is close enough to give a unique starting solution), and to fit not
just arrival times, but also frequencies.  In more detail, using the
long \cxo\ observation as a reference, we first calculated the number
of cycles to the first \xmm\ observation for the best-fit frequency
and for $\dot\nu=0$.  Next, we estimated an uncertainty on the cycle
count using a $\pm5\sigma$ range, with $\sigma$ determined from the
measurement uncertainty on~$\nu$ and an assumed {\em a priori}
uncertainty of $2\times10^{-14}{\rm\,Hz\,s^{-1}}$ on~$\dot\nu$.  For
each possible cycle count, we calculated a new, much more precise
estimate of the frequency, and used it to estimate possible cycle
counts for the next observation.  We iterated this process, deriving
$\dot\nu$ as part of the fit for later iterations, until the fit
clearly became bad ($\chi^2>85$, where including all observations one
has 19 DOF: 11 arrival times plus 11 frequencies minus three fit
parameters), or until all observations were included.  In the end, the
best fits did not depend on which observation was used as an initial
reference.

\begin{figure}
\includegraphics[width=0.9\hsize]{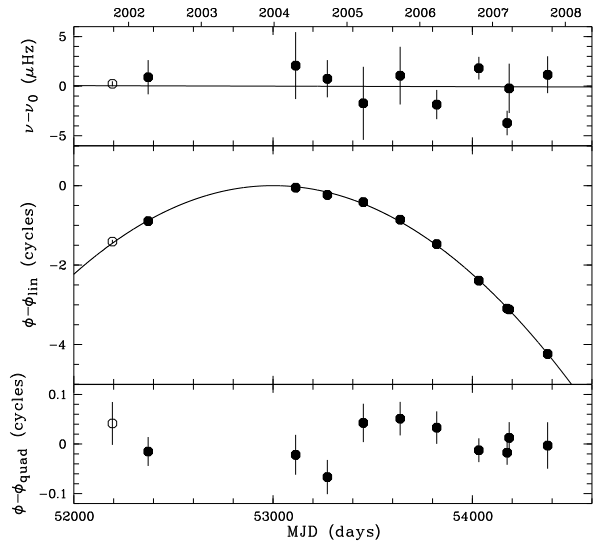}
\caption{Phase and frequency residuals for \rxj.  {\em(top)} Frequency
  residuals relative to the best-fit quadratic model. {\em(center)}
  Phase residuals relative to a linear ($\dot\nu=0$) model.
  {\em(bottom)} Phase residuals relative to the best-fit quadratic
  model.  The filled circles represent \xmm\ observations, and the
  open one the long \chandra\ observation.\label{fig:toa}}
\end{figure}

\begin{deluxetable}{lcc}
\tablewidth{0pt}
\tablecaption{Timing Parameters for \rxj\label{tab:solution}}
\tablehead{& \colhead{Incoherent} & \colhead{Coherent}\\
\colhead{Quantity}  & \colhead{Value} & \colhead{Value}
}
\startdata
Dates (MJD)\dotfill &\multicolumn{2}{c}{52,194--54,377}\\ 
$t_0$ (MJD)\dotfill & 53000.0 & 53000.000009(3)\\ 
$\nu$ (Hz)\dotfill & 0.1417393(6) & 0.1417393685(5)\\ 
$\dot\nu$ ($10^{-16}{\rm~Hz~s^{-1}}$)$\ldots$ & $-50(80)$ & $-5.98(14)$\\ 
TOA rms (s)\dotfill & \nodata & 0.24\\ 
$\chi^2$/DOF\dotfill & 13.7/9 & 25.7/19\\ 
P (s)\dotfill & 7.05521(3) & 7.05520288(2)\\ 
$\dot P$ ($10^{-14}{\rm~s~s^{-1}}$)\dotfill & 20(40) & 2.97(7)\\ 
$\dot E$ ($10^{30}{\rm~erg~s^{-1}}$)\dotfill & $<\!70$ & 3.3\\ 
$B_{\rm dip}$ ($10^{13}$~G)\dotfill & $<\!7$ & 1.5\\ 
$\tau_{\rm char}$ (Myr)\dotfill & $>\!0.17$ & 3.8\\ 
\enddata
\tablecomments{$\tau_{\rm char}=P/2{\dot P}$ is the characteristic
  age, assuming an initial spin period $P_0\ll P$ and a constant
  magnetic field; $B_{\rm dip}=\expnt{3.2}{19}\sqrt{P{\dot P}}$ is the
  magnetic field inferred assuming spin-down by dipole radiation;
  $\dot E=10^{45}I_{45}4\pi^2\nu\dot\nu$ is the spin-down luminosity
  (with $I=10^{45}I_{45}\,\,{\rm g}\,\,{\rm cm}^{2}$ the moment of
  inertia).  Uncertainties quoted are twice the formal 1-$\sigma$
  uncertainties.  For the incoherent analysis, we used the
  $2\,\sigma$ lower limit of $|\dot\nu| <
  1.3\times10^{-14}{\rm\,Hz\,s^{-1}}$ to derive $\dot E$, $B$, and
  $\tau_{\rm char}$.}
\end{deluxetable}

The best trial, shown in Fig.~\ref{fig:toa}, resulted in the frequency
and frequency derivative listed in Table~\ref{tab:solution}.  It has
$\chi^2=25.7$ for 19 DOF.  The fit seems reasonably unique: the two
next-best possibilities have $\chi^2=49$ and 53, then eight are found
between $\chi^2=60$ and~70, and thirteen between 70 and~80.  Also
ignoring frequency information, the best trial is superior: it has
$\chi_{\rm TOA}^2=10.7$ for 8 DOF, while the next best has $\chi_{\rm
  TOA}^2=24$.

As a second method of finding a coherent solution, we calculated
$Z_1^2(\nu,\dot\nu)$ for all data, coherently combining fourier
spectra for individual observations for a range of frequency
derivatives \citep{rem02}.  This has the advantage that we do not have
to decide {\em a priori} which peaks in the individual power spectra
are the correct ones (relevant especially for the long
\cxo\ observation; see Fig.~\ref{fig:ft}).  For the combined data, the
highest peak has $Z_1^2=263$ and occurs at the same $(\nu,\dot\nu)$
found above.  This corresponds to a fractional amplitude
$a=(2Z_1^2/N_{\rm tot})^{1/2}/(1-f_{\rm b})=0.94$\%, consistent with
what was found from the individual observations (here, the total
number of events $N_{\rm tot}=6188356$ and the background fraction
$f_{\rm b}=1.7$\%).  The second and third-highest peaks have
$Z_1^2=234$ and 230, and correspond to the second and ninth-best
solution found using the trial-and-error method, with $\chi^2=49$ and
66.  The changes in ordering arise because here we included the short
observations, which causes the power in the main peak to increase by
$\Delta Z_1^2=5$, but that in most other peaks to decrease (see
Fig.~\ref{fig:ft}).  Folding the shorter observations on the different
solutions confirms this conclusion.

In order to verify the uniqueness of our solution, we ran 1000
simulations in which we assumed our best solution and created photon
time series corresponding to each of the observations (assuming
$a=0.96(1-f_{\rm b})$\%).  We analyzed these in exactly the same way
as the real observations.  Among the simulations, in 983 out of 1000
cases the correct solution was recovered by the trial-and-error method
on the long observations,
and in all 1000 using the $Z_1^2$ power spectra on all data.
Inspection of the mis-identification shows that, as expected,
it is the addition of the information from the shorter observations
that causes the $Z_1^2$ method to do better.

The best-fit reduced $\chi^2$ slightly exceeds unity, with
$\chi^2/{\rm DOF}=1.4$ (the second best solution has $\chi^2/{\rm
  DOF}=2.6$).  This could indicate a fundamental problem, but perhaps
more likely reflects that for low-significance detections, outliers in
phase and frequency happen more often than expected based on a normal
distribution.  Indeed, among our 1000 simulations, we find that 211
have best solutions with $\chi^2>25.7$, somewhat more than the 140
expected for normal distributions.  Alternatively, some unmodeled
phase variations may be present, such as seen in other INS
(\citetalias{kvk05},b; \citealt{vkkpm07}).

\section{Ramifications}\label{sec:discussion}
Assuming the star is spinning down by magnetic dipole radiation, one
can use the spin-down rate to infer a magnetic field strength,
characteristic age, and spin-down luminosity (see
Table~\ref{tab:solution}).  We discuss the ramifications below,
assuming the coherent solution is the correct one.  We compare the
results with those obtained for RX J0720.4$-$3125 and RX J1308.6+2127
(J0720 and J1308 hereafter).

The inferred value of the magnetic field strength of
$\expnt{1.5}{13}\,$G is similar to, but somewhat lower than the values
of 2.4 and $\expnt{3.4}{13}{\rm\,G}$ found for J0720 and J1308.
This lower value might be consistent with the idea that the X-ray
absorption features found in the other sources---but not in
J1856---are due to proton cyclotron lines or transitions in neutral
hydrogen, and that in the lower field of J1856 these are shifted out
of the observed band.  The inferred value, however, is still somewhat
high: calculations by \citet{hkc+07} suggest that for the approximate
temperature of J1856 and $B=\expnt{1.5}{13}\,$G (and for a
gravitational redshift $z_{\rm GR}\simeq0.3$), strong features due to
bound-bound transitions should appear at energies of $\sim\!130\,$eV
(quantum number $m=0\rightarrow1$) and $230\,$eV ($m=0\rightarrow2$),
but none are observed.  One could appeal to elements other than
hydrogen, but these generally have more bound transitions, thus
exacerbating the situation \citep[e.g.,][]{pwl+02,mh07}.

On the other hand, the inferred magnetic field may be consistent with
the idea that the surface is condensed, and that this causes the
black-body like spectrum.  This depends on the composition: from the
recent work by \citet{ml07b}, it appears that for relatively light
elements (based on calculations for carbon and helium), J1856 is far
too hot for condensation to be possible.  For iron, however, the
condensation temperature is relatively high: $kT_{\rm
  cond}\simeq70\,$eV for a magnetic field of $\expnt{1.5}{13}\,$G.
While we do not know the exact surface temperature of J1856 because of
uncertainties in redshift and color correction, the fits of
\citet{hkc+07} have $kT_{\infty}\simeq40\,$eV.  This corresponds to
$kT\simeq55\,$eV at the surface, suggesting that a condensed iron
surface is possible.  Indeed, this might also be the reason that in
early observations, J0720 had a featureless spectrum as well
\citep{pmm+01}.  For its field of $\expnt{2.4}{13}\,$G, iron could
condense below $kT_{\rm cond}\simeq110$\,eV, and its observed
temperature $kT_\infty\simeq85\,$eV corresponds to a surface temperature
of $\sim\!110\,$eV.

The characteristic age of 4\,Myr we derive for J1856 is much larger
than the kinematic age of $0.4{\rm\,Myr}$ inferred assuming an origin
in the Upper Scorpius OB association \citep{wal01,vkk07}, and also
greatly exceeds simple estimates of the cooling age ($\sim 0.5\,$Myr;
e.g., \citealt{pgw06}).  Longer characteristic ages---although by a factor
of three rather than ten---were also found for J0720 and J1308, which
strengthens the suggestion that this is a property common to all
isolated neutron stars (\citetalias{kvk05b}; see \citetalias{kvk05} for
  a discussion of possible causes).

Like for J0720 and J1308, the implied spin-down luminosity $\dot
E\simeq3\times10^{30}{\rm\,erg\,s^{-1}}$ is much smaller than the
X-ray luminosity $L_{\rm X}\simeq\expnt{3}{32}{\rm\,erg\,s^{-1}}$ (for
a distance of $160\,$pc; \citealt{vkk07}), consistent with the lack of
non-thermal emission.  It is also, however, orders of magnitude lower
than the independent estimate of $\dot E\gtrsim
1\times10^{33}(d/160{\rm\,pc})^3{\rm\,erg\,s^{-1}}$ made by assuming
that the H$\alpha$ nebula associated with J1856 is due to a bow shock,
where the pressure from the pulsar wind matches the ram pressure from
the interstellar medium \citep{vkk01b}.  Indeed, this discrepancy
remains even if one considers just the incoherent analysis.  The
alternate model for the H$\alpha$ nebula considered by
\citet{vkk01b}---that it was a moving ionization nebula
\citep{bwm95}---was already rejected by \citet{kvka02}, because the
opening angle of the nebula's tail did not match observations for
distances greater than 100\,pc.  Thus, our new measurement leaves the
nature of the nebula an enigma.

\acknowledgements We thank Marc Cropper for convincing us some years
ago that, in principle, timing solutions could be derived from sparse
data such as ours, and Matthias Ehle and others from the XMM help desk
for investigating ODF problems.  This work made extensive use of the
\xmm\ and \cxo\ archives, as well as ADS, and was supported by NSERC
(MHvK) and NASA (DLK, grant \#01207.01-A).


\begin{thebibliography}{20}
\expandafter\ifx\csname natexlab\endcsname\relax\def\natexlab#1{#1}\fi

\bibitem[{{Blaes} {et~al.}(1995){Blaes}, {Warren}, \& {Madau}}]{bwm95}
{Blaes}, O., {Warren}, O., \& {Madau}, P. 1995, \apj, 454, 370

\bibitem[{{Buccheri} {et~al.}(1983)}]{bbb+83}
{Buccheri}, R. {et~al.} 1983, \aap, 128, 245

\bibitem[{{Cash}(1979)}]{cash79}
{Cash}, W. 1979, \apj, 228, 939

\bibitem[{{Haberl}(2007)}]{haberl07}
{Haberl}, F. 2007, \apss, 308, 181

\bibitem[{{Ho} {et~al.}(2007){Ho}, {Kaplan}, {Chang}, {van Adelsberg}, \&
  {Potekhin}}]{hkc+07}
{Ho}, W.~C.~G., {Kaplan}, D.~L., {Chang}, P., {van Adelsberg}, M., \&
  {Potekhin}, A.~Y. 2007, \mnras, 375, 821

\bibitem[{{Kaplan} \& {van Kerkwijk}(2005{\natexlab{a}})}]{kvk05}
{Kaplan}, D.~L. \& {van Kerkwijk}, M.~H. 2005{\natexlab{a}}, \apjl, 628, L45
  (KvK05a)

\bibitem[{{Kaplan} \& {van Kerkwijk}(2005{\natexlab{b}})}]{kvk05b}
---. 2005{\natexlab{b}}, \apjl, 635, L65 (KvK05b)

\bibitem[{{Kaplan} {et~al.}(2002){Kaplan}, {van Kerkwijk}, \&
  {Anderson}}]{kvka02}
{Kaplan}, D.~L., {van Kerkwijk}, M.~H., \& {Anderson}, J. 2002, \apj, 571, 447

\bibitem[{{Medin} \& {Lai}(2007)}]{ml07b}
{Medin}, Z. \& {Lai}, D. 2007, \mnras, submitted, (arXiv:0708.3863)

\bibitem[{{Mori} \& {Ho}(2007)}]{mh07}
{Mori}, K. \& {Ho}, W.~C.~G. 2007, \mnras, 377, 905

\bibitem[{{Motch} {et~al.}(2003){Motch}, {Zavlin}, \& {Haberl}}]{mzh03}
{Motch}, C., {Zavlin}, V.~E., \& {Haberl}, F. 2003, \aap, 408, 323

\bibitem[{{Paerels} {et~al.}(2001)}]{pmm+01}
{Paerels}, F. {et~al.} 2001, \aap, 365, L298

\bibitem[{{Page} {et~al.}(2006){Page}, {Geppert}, \& {Weber}}]{pgw06}
{Page}, D., {Geppert}, U., \& {Weber}, F. 2006, Nuclear Physics A, 777, 497

\bibitem[{{Pons} {et~al.}(2002)}]{pwl+02}
{Pons}, J.~A. {et~al.} 2002, \apj, 564, 981

\bibitem[{{Ransom} {et~al.}(2002){Ransom}, {Eikenberry}, \&
  {Middleditch}}]{rem02}
{Ransom}, S.~M., {Eikenberry}, S.~S., \& {Middleditch}, J. 2002, \aj, 124, 1788

\bibitem[{{Tiengo} \& {Mereghetti}(2007)}]{tm07}
{Tiengo}, A. \& {Mereghetti}, S. 2007, \apjl, 657, L101 (TM07)

\bibitem[{{van Kerkwijk} \& {Kaplan}(2007)}]{vkk07}
{van Kerkwijk}, M.~H. \& {Kaplan}, D.~L. 2007, \apss, 308, 191

\bibitem[{{van Kerkwijk} {et~al.}(2007){van Kerkwijk}, {Kaplan}, {Pavlov}, \&
  {Mori}}]{vkkpm07}
{van Kerkwijk}, M.~H., {Kaplan}, D.~L., {Pavlov}, G.~G., \& {Mori}, K. 2007,
  \apjl, 659, L149

\bibitem[{{van Kerkwijk} \& {Kulkarni}(2001)}]{vkk01b}
{van Kerkwijk}, M.~H. \& {Kulkarni}, S.~R. 2001, \aap, 380, 221

\bibitem[{{Walter}(2001)}]{wal01}
{Walter}, F.~M. 2001, \apj, 549, 433

\end{thebibliography}
\end{document}